\documentclass[twocolumn,prl,showpacs,amsmath,amssymb]{revtex4}

\usepackage{graphicx}

\def\be{\begin{equation}}
\def\ee{\end{equation}}
\def\bea{\begin{eqnarray}}
\def\eea{\end{eqnarray}}
\def\bma{\begin{mathletters}}
\def\ema{\end{mathletters}}

\def\0{\overline{0}}

\def\q0{\underline{0}}

\def\H{{\cal H}}

\def\Q{{\cal Q}}

\def\L{{\cal L}}

\def\tr{\mbox{tr}}
\def\one{\leavevmode\hbox{\small1\normalsize\kern-.33em1}}

\def\bra#1{\langle#1|} \def\ket#1{|#1\rangle}

\def\proj#1{\ket{#1}\!\bra{#1}}

\begin{document}

\title{ Optimal Bell tests do not require maximally entangled
states
 }

\author{Antonio Ac\'\i n$^{1}$, Richard Gill$^2$ and
Nicolas Gisin$^3$}

\affiliation{ $^1$ICFO-Institut de Ci\`encies Fot\`oniques,
E-08034 Barcelona, Spain\\
$^2$Mathematical Institute, University of Utrecht, Box 80010, 3508
TA Utrecht, The Netherlands\\
$^3$GAP-Optique, University of
Geneva, 20, Rue de l'\'Ecole de M\'edecine, CH-1211 Geneva 4,
Switzerland
 }

\date{\today}


\begin{abstract}
Any Bell test consists of a sequence of measurements on a quantum
state in space-like separated regions. Thus, a state is better
than others for a Bell test when, for the optimal measurements and
the same number of trials, the probability of existence of a local
model for the observed outcomes is smaller. The maximization over
states and measurements defines the optimal nonlocality proof.
Numerical results show that the required optimal state does not
have to be maximally entangled.
\end{abstract}


\maketitle

As first shown by Bell \cite{Bell} in 1964, the correlations among
the measurement outcomes of space-like separated parties on some
quantum states cannot be reproduced by a local theory. This fact
is often referred to as quantum nonlocality and has been
recognized as the most intriguing quantum feature. The fundamental
importance of the work by Bell was that it provided conditions for
experimentally testing Quantum Mechanics (QM) versus the whole set
of local models, the so-called Bell inequalities. The experimental
demonstration \cite{exp}, up to some loopholes, of a Bell
inequality violation definitely closed the Einstein-Podolsky-Rosen
program \cite{EPR} for the existence of a local theory alternative
to QM.

The interest on quantum correlations, or entanglement, 
has dramatically increased during the last two decades due to the
emerging field of Quantum Information Science (QIS) \cite{book}.
It has been realized that quantum states provide new ways of
information processing and communication without analog in
Classical Information. The essential resource for most of these
applications are entangled states. This has motivated a strong
effort devoted to the characterization and quantification of the
entanglement of quantum states. Although many questions remain
unanswered, the problem is completely solved for the case of pure
states in bipartite systems. For a state
$\ket{\Psi}\in\H_A\otimes\H_B$, its amount of entanglement is
specified by the so-called entropy of entanglement \cite{BBPS},
$E(\Psi)=S(\rho_A)$, where $S$ is the usual von Neumann entropy
and $\rho_A=\tr_B(\proj{\Psi})$. In particular, this means that
the maximally entangled state in a bipartite system of dimension
$d\times d$, reads
\begin{equation}\label{maxent}
    \ket{\Phi_d}=\frac{1}{\sqrt d}\sum_{i=1}^d\ket{ii} ,
\end{equation}
where $\{i\}$ define orthonormal bases in $\H_A$ and $\H_B$.

Apart from their importance for quantum information applications,
entangled states provide the only known way of establishing
nonlocal correlations among space-like separated parties. It is
meant here by nonlocal those correlations that (i) cannot be
explained by a local model but (ii) do not allow any
faster-than-light communication, that is, they are consistent with
the no-signaling condition. Indeed, it is a well-established
result that a quantum pure state violates a Bell inequality if and
only if it is entangled \cite{Gisin}. However, it is also well
known that there exist nonlocal correlations that are not
achievable by measuring quantum states \cite{PR}. In a similar
line of thought, it has very often been assumed that
$\ket{\Phi_d}$ represents the most nonlocal quantum state too.
However, no precise demonstration of this fact has ever been given
and, indeed, it is one of the scopes of this work to raise some
doubts about this statement.

In what follows, entangled states constitute a resource for
constructing nonlocality proofs. The strength of a Bell experiment
has to be computed by means of statistical tools: a Bell test is
better than another when, for the same number of trials, the
probability that a local model explains the observed outcomes is
smaller. Recall that statistical fluctuations on finite samples
allow a local theory to predict the possibility of data violating
a Bell inequality. The goal is then to identify those states
needed in the construction of optimal Bell tests. The importance
of constructing optimal nonlocality proofs is two-fold. First,
from an experimental point of view, they allow improving present
Bell experiments, especially in terms of the needed resources.
Second, Bell tests also represent an important tool for QIS
\cite{useful}. In particular, they are useful for testing the
quantumness of devices. This is a hardly explored problem in QIS
that, for instance, can be especially relevant in cryptographic
applications \cite{MY}: given some observed correlations among
several parties, how can its quantum origin be certified? Could
these correlations have alternatively been established by
classical means, i.e. shared randomness? Bell inequalities provide
an answer to the previous questions.


{\sl The scenario:} We consider the standard scenario for any Bell
test. Two space-like separated parties, called Alice and Bob,
share copies of a pure quantum state $\ket{\Psi}\in
\H_A\otimes\H_B$, of dimension $d\times d$. They can choose among
$m$ possible measurements, each of $n$ outcomes, this being
denoted by $m\times n$. $A_j^i$ denotes the positive operator
corresponding to the outcome $j$ of the measurement $i$ for Alice,
so $\sum_j A_j^i=\one$. Similarly, Bob's measurement operators are
denoted by $B_j^i$. The probability that Alice and Bob obtain the
outcomes $j_A$ and $j_B$ after applying the measurement $i_A$ and
$i_B$, where $j_A,j_B=1,\ldots,n$ and $i_A,i_B=1,\ldots,m$, on
$\ket{\Psi}$ is
\begin{equation}\label{prob}
    p_Q(j_A,j_B|i_A,i_B)=\tr\left(A_{j_A}^{i_A}\otimes
    B_{j_B}^{i_B}\proj{\Psi}\right) .
\end{equation}
In a Bell experiment, a quantum state is prepared and sent to the
parties who measure it. After $N$ repetitions of the experiment,
the frequencies of the results define a $(m^2n^2)$-dimensional
vector whose components tend to $p_Q(j_A,j_B|i_A,i_B)$ when
$N\rightarrow\infty$. A vector of probabilities is achievable
using QM when there exist a state $\ket{\Psi}$ and measurements
$A_{j_A}^{i_A}$ and $B_{j_B}^{i_B}$ satisfying (\ref{prob}).

On the other hand, in a local model any observed correlation
between measurement results in space-like separated regions should
come from initially shared random data, denoted by $\lambda$. QM
is nonlocal because some of the vectors (\ref{prob}) do not allow
a local description, i.e. they cannot be written as
\begin{equation}\label{prloc}
    p_L(j_A,j_B|i_A,i_B)=\sum_{\lambda}p\,(\lambda)
    p\,(j_A|i_A,\lambda)p\,(j_B|i_B,\lambda) .
\end{equation}
Therefore, shared quantum states can be used to establish nonlocal
correlations.

The goal of any Bell experiment is to test the hypothesis $\Q$,
``the observed outcomes are governed by a quantum probability
distribution (\ref{prob})", against the composite hypothesis $\L$,
``there exists a local model (\ref{prloc}) reproducing the data"
\cite{vDGG}. The statistical tool that quantifies the {\sl average
amount of support in favor of $\Q$ against $\L$ per trial when the
data are generated by $\Q$} is the so-called relative entropy or
Kullback-Leibler (KL) Divergence \cite{CT}, $D$. For two
probability distributions, $\vec p_1$ and $\vec p_2$ associated to
the event $\{z\}$, it reads
\begin{equation}\label{relent}
    D(\vec p_1||\vec p_2)=\sum_z
p_1(z)\log(p_1(z)/p_2(z)) .
\end{equation}
We denote by $\vec q=(q(1,1,1,1),\ldots,q(m,m,n,n))$ the quantum
probabilities for the measurement settings $i_A$ and $i_B$, and
outcomes $j_A$ and $j_B$. Using (\ref{prob}), one has
\begin{equation}\label{qprob}
    q(i_A,i_B,j_A,j_B)=\tr\left(A_{j_A}^{i_A}\otimes
    B_{j_B}^{i_B}\proj{\Psi}\right)p_M(i_A,i_B) ,
\end{equation}
where $p_M(i_A,i_B)$ characterizes the choice of measurements by
Alice and Bob. Now, the support in favor of $\Q$ against $\L$
provided by these quantum data is \cite{vDGG}
\begin{equation}\label{KLent}
    D(\Q||\L)=\min_{\vec p\in\L}D(\vec q\,||\vec p) ,
\end{equation}
where the minimization runs over all alternative local models,
$\vec p$. The vector $\vec p$ is defined analogously to
(\ref{qprob}), replacing the quantum term $p_Q$ (\ref{prob}) by a
local model $p_L$ (\ref{prloc}). This quantity gives the
statistical figure of merit to be maximized in any nonlocality
test \cite{vDGG}. It is worth mentioning here that the KL
Divergence (i) is an asymptotic measure and (ii) appears as the
measure of statistical support for the two most commonly used
methods for hypothesis testing, frequentist and Bayesian (see
\cite{vDGG} for more details). Moreover, and despite not being
symmetric, it can be seen as a measure of statistical distance
between probability distributions.

It is often convenient to interpret a Bell test as a game between
a quantum and a local player \cite{vDGG,Peres}. The quantum player
has to design an experimental situation for which the local player
is unable to provide a model. Thus, the quantum player looks for
the experiment that gives him the victory with the minimal number
of repetitions, i.e. his task consists of designing the Bell test
maximizing (\ref{KLent}). In order to do that, he can choose the
state to be prepared, the measurements, and the probability
governing the choice of measurements, $p_M(i_A,i_B)$. Notice that
we do not impose $p_M(i_A,i_B)$ to be product. Indeed, one could
think of a configuration where an external referee is sending the
choice of measurements to the parties. On the other hand, the
local player only assumes the existence of a local model. In
particular, he is allowed to change his description according to
the observed data.


{\sl Results:} In what follows, the optimality of Bell tests is
analyzed according to the KL Divergence. The optimization of
(\ref{KLent}) in full generality is a very hard problem. Here, we
mainly consider the standard situation where Alice and Bob apply
two projective measurements, i.e. $m=2$, $n=d$ and $A_i^j$ and
$B_i^j$ are mutually orthogonal one-dimensional projectors. For a
fixed number of measurements, it is possible to search numerically
the state and measurements defining the optimal Bell test. In the
qubit case, $d=2$, the best nonlocality proof is given by the
maximally entangled state $\ket{\Phi_2}$ (\ref{maxent}) and the
measurements maximizing the violation of the CHSH inequality
\cite{CHSH}, as expected. The KL Divergence turns out to be equal
to 0.046 bits \cite{vDGG} and the optimal choice of settings is
completely random, $p_M(i_A,i_B)=1/4$. Actually, the optimal
choice of settings turns out to be random for all the situations
considered in this work.

Moving to higher dimension, the optimal measurements for the
maximally entangled state of two three-dimensional systems are the
ones maximizing its violation of the CGLMP inequality
\cite{CGLMP}. The statistical strength is of 0.058 bits,
reflecting the fact that quantum nonlocality increases with the
dimension \cite{KGZMZ}. However, it is known that the largest
violation of the CGLMP inequality is given by a nonmaximally
entangled state \cite{ADGL}
\begin{equation}\label{mvstate}
    \ket{\Psi^{mv}_3}=
    \gamma(\ket{00}+\ket{11})+\sqrt{1-2\gamma^2}\ket{22} ,
\end{equation}
where $\gamma\approx 0.617$. The measurements maximizing its
statistical strength are again those maximizing its Bell violation
(which are the same as for $\ket{\Phi_3}$) and give 0.072 bits,
larger than the value obtained for the maximally entangled state.
The maximization now over the space of measurements, choices of
settings and states gives the same measurements as above but for a
different state,
\begin{equation}\label{nlstate}
    \ket{\Psi_3}\approx
    \delta(\ket{00}+\ket{11})+\sqrt{1-2\delta^2}\ket{22} ,
\end{equation}
where $\delta\approx 0.642$. Therefore, the $3\times 3$ state
producing the optimal nonlocality test with two projective
measurements per site does not have maximal entanglement.
Actually, this state has even less entanglement than
$\ket{\Psi^{mv}_3}$. All these results are summarized in Table I.
It is worth mentioning here that the optimal measurements are the
same for all three states. Similar results are obtained for $d=4$:
(i) the optimal measurements are those maximizing the Bell
violation for $\ket{\Phi_4}$ \cite{CGLMP,ADGL} but (ii) the
optimal state, $\ket{\Psi_4}$, is not maximally entangled. The
corresponding KL Divergence is of 0.098 bits. The problem in full
generality becomes intractable for larger $d$, so the following
simplifications are considered.

\begin{table}\label{qutrsumm}
\begin{tabular}{|c|c|c|}
    \hline State & KL Divergence (bits) & Entanglement (bits) \\
\hline \hline
  $\ket{\Phi_3}$ & 0.058 & 1.585 \\
  $\ket{\Psi^{mv}_3}$ & 0.072 & 1.554 \\
  $\ket{\Psi_3}$ & 0.077 & 1.495 \\
\hline
\end{tabular}
\caption{Nonlocal content and entanglement for the maximally
entangled state, $\ket{\Phi_3}$, the state maximally violating the
CGLMP inequality, $\ket{\Psi^{mv}_3}$, and the optimal state
$\ket{\Psi_3}$.}
\end{table}

First, the $2\times d$ measurements are taken equal to those
maximizing the Bell violation for $\ket{\Phi_d}$: the parties
apply a unitary operation with only nonzero terms in the diagonal,
$e^{i\phi_a(j)}$ for Alice and $e^{i\varphi_b(j)}$ for Bob, with
$j=0,\ldots,d-1$ and $a,b=1,2$. These phases read \cite{CGLMP}
\begin{equation}
\label{phase} \phi_1(j)=0 \quad\phi_2(j)=\frac{\pi}{d}j \quad
\varphi_1(j)=\frac{\pi}{2d}j \quad \varphi_2(j)=-\frac{\pi}{2d}j .
\end{equation}
Then, Alice carries out a discrete Fourier transform, $U_{FT}$,
and Bob applies $U_{FT}^*$, and they measure in the computational
basis. Thus, it is assumed in what follows that these measurements
define the optimal $2\times d$ Bell test. This is known to be the
case for qubits, and our numerical results indicate that this also
happens for $d=3,4$.

Once the settings are fixed, the problem is cast in a formulation
very similar to a standard Bell inequality. The goal is now to
obtain the $d\times d$ state maximizing (\ref{KLent}) for the
given settings. Let $\vec q\,^s$ and $\vec p\,^s$ denote a pair
forming a solution to this problem, i.e.
\begin{equation}
\label{klsol}
    \max_{\vec q}\,\min_{\vec p}D(\vec q\,||\vec p)=D(\vec q\,^s\,||
    \vec p\,^s)=D^s .
\end{equation}
For small deviations from this solution one has $D(\vec
q\,^s+\delta\vec q\,||\vec p\,^s)\leq D^s$ and $D(\vec
q\,^s\,||\vec p\,^s+\delta\vec p)\geq D^s$. Therefore, all vectors
of quantum and local probabilities close to the previous solution
satisfy
\begin{eqnarray} \label{bellop}
  \sum_i \log\left(\frac{q^s_i}{p^s_i}\right) q_i  &\leq& D^s \\
\label{bellin}
  \sum_i \frac{q^s_i}{p^s_i} p_i &\leq& 1 .
\end{eqnarray}
The values $D^s$ and 1 are found by substituting $q_i=q^s_i$ and
$p_i=p^s_i$, respectively. Actually this has to be true for all
$\vec q$ and $\vec p$. If this was not the case, using convexity
arguments one could construct a point arbitrarily close to $\vec
p\,^s$ or $\vec q\,^s$ violating these conditions. Indeed, assume
there exists a vector of quantum probabilities $\vec q'$ violating
(\ref{bellop}). Then, $(1-\epsilon)\vec q^s+\epsilon\,\vec q'$
would also violate the same condition for arbitrarily small
$\epsilon$. Note that the quantity on the left hand side of
(\ref{bellop}) can be seen as the mean value of a Bell operator,
while (\ref{bellin}) defines a Bell inequality. Then, $\vec q\,^s$
maximizes (\ref{bellop}) over all $\vec q$, while $\vec p\,^s$
does it for (\ref{bellin}).

After inspection, one can see that the Bell inequality
(\ref{bellin}) corresponding to the optimal solution for $d=2,3,4$
is of the CGLMP form, up to taking a linear combination with the
normalization condition $\sum_i p_i=1$. Actually, Eq.
(\ref{bellin}) can be rewritten in these three cases as
\begin{equation}\label{gillin}
    \langle[A_1-B_1]+[B_1-A_2]+[A_2-B_2]+[B_2-A_1-1]\rangle\geq d-1 ,
\end{equation}
where $[X]$ stands for $X$ modulo $d$ and $\langle
X\rangle=p\,(X=1)+ 2p\,(X=2)+\ldots+(d-1)p\,(X=d-1)$. This
inequality easily follows from the identity
\begin{equation}
    [A_1-B_1  +  B_1-A_2  +  A_2-B_2   +  B_2-A_1-1]  =  d-1 ,
\end{equation}
and the fact that $[X]+[Y]\geq[X+Y]$. One can see that Eq.
(\ref{gillin}) represents an extremely compact way of writing all
CGLMP inequalities for arbitrary dimension. Then, it is assumed
that the inequality (\ref{bellin}), derived from Eq.
(\ref{klsol}), has the CGLMP form (\ref{gillin}), up to linear
combination with $\sum_i p_i=1$, also for $d>4$. Thus the
$q_i^s/p_i^s$ terms in (\ref{bellop}) and (\ref{bellin}) are known
functions of one parameter.

\begin{figure}
\begin{center}
  \includegraphics[width=8 cm]{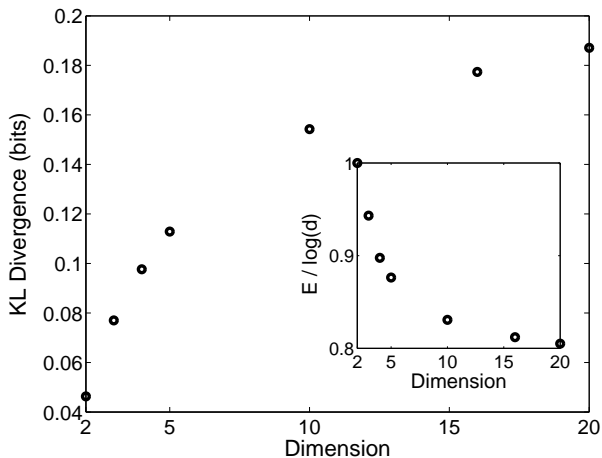}\\
  \caption{KL Divergence for Bell tests using two projective
  measurements under the mentioned assumptions. In the inset, it is shown
  the amount of entanglement, in terms of Entropy of Entanglement,
  for the optimal state, $\ket{\Psi_d}$.}\label{optst}
\end{center}
\end{figure}

The problem has now been hugely simplified. Under the mentioned
assumptions, the state for an optimal $2\times d$ Bell test is
given by the eigenvector of largest eigenvalue of the Bell
operator (\ref{bellop}), where the measurements are fixed as
before, and where the coefficients of the Bell operator are
determined (up to one unknown parameter, over which we also
optimize) by the CGLMP inequality. The associated eigenvalue gives
the optimal KL Divergence. This computation can be done up to very
large dimension, the results can be found in Fig.
\ref{optst}. 

{\sl Discussion:} Figure \ref{optst} shows several interesting
features. First of all, one can see that for an optimal $2\times
d$ Bell test, there is no need for systems of very large
dimension. Actually, the simplest CHSH scenario for the singlet
state already constitutes a reasonably good test for ruling out
local models. However, beyond this simple case, none of the
optimal Bell tests requires a maximally entangled state. In all
the studied situations, the Schmidt basis for the optimal state
was the computational one. Assuming this is always the case, we
can compute the conjectured optimal state for large $d$, say
$d=1000$, finding that $E(\Psi_d)\rightarrow\ln d\approx 0.69\log
d$ bits. 

It also follows from Fig. 1 that two measurements per site may not
be optimal for large $d$. For instance, when $d=16$ the
conjectured optimal $2\times 16$ test is worse than the $4\times
16$ test consisting of two independent realizations of the optimal
$2\times 4$ Bell test for two copies of $\ket{\Psi_4}$. Indeed, it
is always possible to interpret two independent realizations of
this Bell test as a ``new" $2^2\times 4^2$ Bell test for
$\ket{\Psi}^{\otimes 2}$. Using that (i) the KL Divergence is
additive, $D(\vec q\,^2||\vec p\,^2)=2D(\vec q\,||\vec p)$, and
(ii) the closest local model to two independent realizations of
the same Bell test corresponds to two independent realizations of
the best local model for the single-copy case, the KL Divergence
for this test is twice the initial one.

A priori, one would have expected the maximally entangled state to
be the optimal state for any Bell test. A thorough numerical
search of Bell tests for the maximally entangled state
$\ket{\Phi_3}$ using more settings per site and general
measurements has been performed. No improvement over the
optimal $2\times 3$ case was obtained. 
Actually, it is remarkable that Bell tests with two projective
measurements per site are so good for low dimensional systems.
Therefore, all the previous numerical results show that beyond
qubits and for the same amount of resources (system dimension and
number and type of settings) the optimal state for a Bell test is
not maximally entangled.

{\sl Conclusions:} Non-local correlations constitute an
information theoretic resource {\sl per se} \cite{BLMPPR}, that
can be distributed by means of quantum states. It is known that
there are nonlocal correlations that cannot be established by
measuring quantum states \cite{PR}. Moreover, the nonlocal
correlations obtained from the maximally entangled state
$\ket{\Phi_3}$ seem to be less robust against noise than those
from $\ket{\Psi^{mv}_3}$ \cite{ADGL}. Actually, the communication
cost of simulating the nonlocal correlations for
$\ket{\Psi^{mv}_3}$ seems to be higher than for $\ket{\Phi_3}$
\cite{Pironio}. More recently, it has been shown that the
so-called nonlocal machine \cite{PR,BLMPPR} is sufficient for the
simulation of the correlations in a singlet state \cite{CGMP}, but
it fails for some nonmaximally entangled states of two qubits
\cite{BGS}. All these result suggest that, despite the fact that
all pure entangled states contain nonlocal correlations
\cite{Gisin}, the relation between entanglement and nonlocality is
subtler than firstly expected, since they may represent different
information resources.

In this work, entangled states are analyzed as a tool for the
construction of Bell tests. For all the studied scenarios and
beyond the qubit case, the states needed for an optimal Bell test
are not maximally entangled.


\bigskip

This work is supported by the ESF, an MCYT ``Ram\'on y Cajal"
grant, the Generalitat de Catalunya, the Swiss NCCR ``Quantum
Photonics" and OFES within the EU project RESQ (IST-2001-37559).


\begin{references}

\bibitem{Bell}
J. S. Bell, Physics {\bf 1}, 195 (1964).

\bibitem{exp}
A. Aspect, P. Grangier and G. Roger, Phys. Rev. Lett. {\bf 47},
460 (1981); W. Tittel {\sl et al.}, {\sl ibid} {\bf 81}, 3563
(1998); G. Weihs {\sl et al.}, {\sl ibid} {\bf 81}, 5039 (1998);
M. Rowe {\sl et al.}, Nature {\bf 409}, 791 (2001).

\bibitem{EPR}
A. Einstein, B. Podolsky and N. Rosen, Phys. Rev. {\bf 47}, 777
(1935).


\bibitem{book}
See for instance M. A. Nielsen and I. L. Chuang, {\sl Quantum
Computation and Quantum Information}, Cambridge University Press
(2000).

\bibitem{BBPS}
C. H. Bennett, H. J. Bernstein, S. Popescu and B. Schumacher,
Phys. Rev. A {\bf 53}, 2046 (1996).

\bibitem{Gisin}
N. Gisin, Phys. Lett. A {\bf 154}, 201 (1991).

\bibitem{PR}
S. Popescu, D. Rohrlich, Found. Phys. {\bf 24}, 379 (1994).

\bibitem{useful}
A. Ac\'\i n, N. Gisin, L. Masanes and V. Scarani,  Int. J. Quant.
Inf. {\bf 2}, 23 (2004).

\bibitem{MY}
D. Mayers and A. Yao, Quant. Inf. Comp. {\bf 4}, 273 (2004).

\bibitem{vDGG}
W. van Dam, R. Gill and P. Gr\"unwald, quant-ph/0307125; to appear
in IEEE Trans. Inf. Theory.

\bibitem{CT}
T. M. Cover and J. A. Thomas, {\sl Elements of Information
Theory}, Wiley Interscience, New York (2000).

\bibitem{Peres}
A. Peres, Fortsch. Phys. {\bf 48}, 531 (2000).


\bibitem{CHSH} J. F. Clauser, M. A. Horne, A. Shimony, R. A.
Holt, Phys. Rev. Lett. {\bf 23}, 880 (1969).

\bibitem{CGLMP}
D. Collins {\sl et al.}, Phys. Rev. Lett. {\bf 88}, 040404 (2002).

\bibitem{KGZMZ}
D. Kaszlikowski {\sl et al.}, Phys. Rev. Lett. {\bf 85}, 4418
(2000).

\bibitem{ADGL}
A. Ac\'\i n, T. Durt, N. Gisin and J. I. Latorre, Phys. Rev. A
{\bf 65}, 052325 (2002).

\bibitem{BLMPPR}
J. Barrett {\sl et al.}, Phys. Rev. A {\bf 71}, 022101 (2005).


\bibitem{Pironio}
S. Pironio, Phys. Rev. A {\bf 68}, 062102 (2003).

\bibitem{CGMP}
N. J. Cerf, N. Gisin, S. Massar and S. Popescu, Phys. Rev. Lett.
{\bf 94}, 220403 (2005).

\bibitem{BGS}
N. Brunner, N. Gisin and V. Scarani, New J. Phys. {\bf 7}, 88
(2005).




\end{references}
\end{document}